%% file: RR-8267.tex
\documentclass[a4paper]{article}
\usepackage[latin1]{inputenc} 
\usepackage[T1]{fontenc} 
\usepackage{RR,RRthemes}
\usepackage{hyperref}
\RRdate{Mars 2013}

\usepackage{amsfonts}
\usepackage{amsmath}
\usepackage{algorithm} 
\usepackage{algorithmic}
\usepackage{subfigure}
\usepackage{tikz}

\newcommand{\ele}{{\ell}}
\newtheorem{lemma}{Lemma}
\newcommand{\bproof}{\textbf{Proof.}~}
\newcommand{\eproof}{$\spadesuit$\\}

\RRauthor{
Franco Robledo
  \and
Pablo Sartor
}
\authorhead{Robledo \& Sartor}
\RRtitle{Une m\'ethode de simulation pour l'estimation de la performabilit\'e des r\'eseaux utilisant pathsets et cutsets calcul\'es heuristiquement}
\RRetitle{A Simulation Method for Network Performability Estimation using Heuristically-computed Pathsets and Cutsets}
\titlehead{Performability Estimation with Edge-sets Heuristics}
\RRresume{Consid\'erez un ensemble de n{\oe}uds terminaux K appartenant \`a un r\'eseau dont les n{\oe}uds sont reli\'es par des liaisons qui \'echouent ind\'ependamment avec des probabilit\'es connues. Nous pr\'esentons une m\'ethode pour estimer n'importe quelle mesure de performabilit\'e qui d\'epend de la distance en sauts entre les n{\oe}uds terminaux. Elle g\'en\'eralise des m\'ethodes de Monte Carlo pr\'ec\'e-demment introduites pour l'estimation de la K-fiabilit\'e des r\'eseaux avec r\'eduction de la variance par rapport \`a Monte Carlo standard. Ces m\'ethodes sont bas\'ees sur l'utilisation d'ensembles d'ar\^etes design\'es d-pathsets et d-cutsets pour r\'eduire la variance de l'estimateur. Ces ensembles d'ar\^etes, consid\'er\'es comme connus a priori dans la litt\'erature pr\'ec\'edente, affectent fortement les performances atteintes ; nous introduisons et comparons une famille d'heuristiques pour leur s\'election. Des exemples num\'eriques sont pr\'esent\'es, montrant les importantes  am\'eliorations dans l'efficacit\'e qui peuvent \^etre obtenues par le cha\^inage de ces heuristiques avec le plan d'\'echantillonnage de Monte Carlo propos\'e.
}
\RRabstract{Consider a set of terminal nodes K that belong to a network whose nodes are connected by links that fail independently with known probabilities. We introduce a method for estimating any performability measure that depends on the hop distance between terminal nodes. It generalises previously introduced Monte Carlo methods for estimation of the K-reliability of networks with variance reduction compared to crude Monte Carlo. They are based on using sets of edges named d-pathsets and d-cutsets for reducing the variance of the estimator. These sets of edges, considered as a priori known in previous literature, heaviliy affect the attained performance; we hereby introduce and compare a family of heuristics for their selection. Numerical examples are presented, showing the significant efficiency improvements that can be obtained by chaining the edge set selection heuristics to the proposed Monte Carlo sampling plan.}
\RRmotcle{heuristiques, pathset, cutset, Monte Carlo, \'ev\'enements rares, fiabilit\'e, performabilit\'e, longueur born\'ee, restrictions de diam\`etre}
\RRkeyword{heuristics, pathset, cutset, Monte Carlo, rare events, reliability, performability, bounded length, diameter constraints}

\RRprojets{Dionysos}
\RRdomaine{3} 

 \URRennes  

\begin{document}
\RRNo{8267}
 \makeRT 

\section{Introduction}

Consider a communication network whose components randomly fail, modelled as an undirected graph. The most classical reliability analysis model assigns to the network two possible states determined by the link states; it is operational if and only if a certain set of distinguished sites, known as terminals, are connected, otherwise it is failed. A generalization introduced in \cite{PR01} imposes also an upper limit in the allowed distance between terminals for the network to be considered as operational. This generalization, conceived to model situations where limits exist in the acceptable delay times or the number of hops undergone by data packets, keeps the binary nature of the network state. But when in comes to performability, in several contexts there is need to employ metrics defined over a larger number of network states, characterised by the hop distance between terminals. For example, in voice-over-IP applications, the perceived quality is affected by latency, which is in turn determined by the number of links traversed by packets. Quality deteriorates as high hop-distance states occur more frequently in the network. In web applications with rich interfaces, the quality perceived by the end user is related to responsiveness, where latency determines the delay between the user actions and their effect on the output interface. The same consideration applies to other contexts where costs are born each time a link is traversed e.g. vehicle or packet routing, where costs can relate to time spent, tolls, fuel, etc.  

Classical reliability analysis consists of computing, estimating or bounding the probability that the network is operational, that is, the expected value of the binary variable associated with the network state. Surveys can be found in \cite{Colbourn87}, \cite{Rub96} and \cite{Petingi2008}. In particular, many Monte Carlo methods have been proposed for efficient estimation of this expected value. This article introduces a Monte Carlo method to estimate any performability metric that is defined as the expected value of a distance-dependent network metric. In real communication networks, link reliabilities are normally very high. Then, when applying simulation methods, sampling a `high distance' network state is a rare event. In the context of Monte Carlo simulations for network reliability analysis, once fixed a certain confidence interval goal, the needed sample size unboundedly grows as link reliabilities become higher. Several variance-reduction techniques can be used to reduce the sample size; surveys can be found in \cite{rubino2009rare}, \cite{Can-ElKha-Rub-2009} and \cite{Gertsbakh2009}; more recent works include \cite{Can-LEcu-Lee-Rub-Tuf-2009}, \cite{Can-LEcu-Rub-Tuf-2010}, \cite{LEcu-Rub-Sag-Tuf-2011}, \cite{Zenklusen-Laumanns-2011} and \cite{Bot-LEcu-Rub-Sim-Tuf-2012}. \cite{KTI1977} and \cite{Fishman1986} introduced a family of methods for the classical reliability, based on sampling strategies conditioned by paths and cuts. In \cite{SartorCOMCOM2012} it is shown how to extend them to include diameter constraints, employing sets of edges named $d$-pathsets and $d$-cutsets. The Monte Carlo method hereby proposed generalises these methods in order to estimate the expected value of a random variable determined by the maximal distance between pairs of terminals.

The $d$-pathsets and $d$-cutsets, that were considered as \emph{a priori} known in the previously mentioned literature, heaviliy affect the performance attained by the simulation methods. We hereby introduce and compare a family of heuristics for their selection. We present numerical evidence of the significant efficiency gains attained by chaining these heuristics to the proposed simulation method when compared to a crude Monte Carlo simulation. 

The remainder of the article is organised as follows. Section~\ref{s:defin} includes definitions, notation and model formalisation. Section~\ref{s:crude} describes the crude Monte Carlo method and the estimators it gives for the metric under study as well as for its variance. Section~\ref{s:bounded} describes the suggested Monte Carlo method and shows the variance reductions achieved relative to the crude one. Section~\ref{s:heurisPatCut} describes the heuristics for selecting the ($d$-)pathsets and ($d$-)cutsets that will be applied in the simulation. Section~\ref{s:tests} presents three numerical examples, based on mesh-like networks. It compares several variations of the heuristic, the relative efficiency of the proposed Monte Carlo method vs. the crude one and how it is influenced by link reliabilities. Finally, conclusions and further work are summarised in Section~\ref{s:conclusions}.

\section{Definitions and Notation}\label{s:defin}

The network is modelled by an undirected graph $G=(V,E)$ with $n=|V|$, $m=|E|$ and $E=\{e_1...e_m\}$, whose nodes and edges correspond to the sites and links of the network respectively. The following definitions and notation are also employed:

\begin{itemize}
\item $K\subseteq V$: a subset of nodes that corresponds to the distinguished sites (called \emph{terminal nodes} or simply \emph{terminals});
\item $X_e$: for every $e{\in}E$, a random binary variable whose value is $1$ if $e$ operates and $0$ otherwise;
\item $r_e$: reliability of $e$ (the probability that it is operational at any given instant). The edges are assumed to fail independently of one another;
\item $X=(X_1...X_m)\in\{0,1\}^m$: a \emph{network configuration} (an $m$-tuple encoding the states of all edges);
\item $\cal X$: set of the $2^m$ possible network configurations;
\item $\pi(x) = \Pr(X=x)$ (probability that the random network configuration is $x$);
\item $\Delta:{\cal X}\rightarrow \{0,...,n,\infty\}$: the function that gives the maximum distance $\Delta(x)$ between two terminals in the partial graph of $G$ encoded by a network configuration $x$;
\item $K$ is \emph{$d$-connected} in $G$ if and only if, for each pair of nodes of $K$, there is a path between them, whose length is not above the integer $d$;
\item $\Phi$: network parameter to estimate (random variable determined by the network configurations).
\end{itemize}

Our goal is to estimate the expected value of the random variable $\Phi$. The value of this random variable is determined by the network configurations as follows. The set $\{0,...,n,\infty\}$ (the codomain of $\Delta$) is partitioned into several intervals (think of them as `quality levels'). Then each interval is mapped to one value from a set ${\cal Q}\subset {\mathbb R}$ with $|{\cal Q}|\leq r+1$. Therefore every network configuration $x$ belongs to an unique `quality level' that corresponds to a certain $\Phi$ value. Then, our final aim is to estimate the expected value of the function $\Phi: {\cal X}\to {\cal Q}$.

\begin{figure}[h!]\centering{

\begin{tikzpicture} [scale=1.5,nodop/.style={inner sep=0pt
}]
	\path[draw] (3.4,2) -- (1,2) -- (1,4) -- (3.4,4);
	\path[draw] (3.6,2) -- (6,2) -- (6,4) -- (3.6,4); 
	\path[draw] (3.3,1.8) -- (3.5,2.2);
	\path[draw] (3.5,1.8) -- (3.7,2.2);
	\path[draw] (3.3,3.8) -- (3.5,4.2);
	\path[draw] (3.5,3.8) -- (3.7,4.2);

	\draw[dotted] (2,4) -- (2,0);
	\draw[dotted] (3,4) -- (3,0);
	\draw[dotted] (4,4) -- (4,0);
	\draw[dotted] (5,4) -- (5,0);

	\node[nodop] at (0.5,1.3) {$\Delta$};
	\node[nodop] at (1.5,1.3) {$(0,d_0]$};
	\node[nodop] at (2.5,1.3) {$(d_0,d_1]$};
	\node[nodop] at (4.5,1.5) {$(d_{r-2},$};
	\node[nodop] at (4.5,1.2) {$d_{r-1}]$};
	\node[nodop] at (5.5,1.5) {$(d_{r-1},$};
	\node[nodop] at (5.5,1.2) {$\infty]$};

	\node[nodop] at (0.5,0.7) {$regions$};
	\node[nodop] at (1.5,0.7) {${\cal X}_0$};
	\node[nodop] at (2.5,0.7) {${\cal X}_1$};
	\node[nodop] at (4.5,0.7) {${\cal X}_{r-1}$};
	\node[nodop] at (5.5,0.7) {${\cal X}_r$};

	\node[nodop] at (0.5,0.2) {$\Phi$};
	\node[nodop] at (1.5,0.2) {$\Phi_0$};
	\node[nodop] at (2.5,0.2) {$\Phi_1$};
	\node[nodop] at (4.5,0.2) {$\Phi_{r-1}$};
	\node[nodop] at (5.5,0.2) {$\Phi_r$};

	\node at (0.5,3) {$\cal X$};

	\draw[style=dotted,thick] (1.5,3.1) ellipse (.4 and .6) node {$z_0$};
	\draw[style=dotted,thick] (2.5,3.1) ellipse (.4 and .6) node {$z_1$};
	\draw[style=dotted,thick] (4.5,3.1) ellipse (.4 and .6) node {$z_{r-1}$};
	\draw[style=dotted,thick] (5.5,3.1) ellipse (.4 and .6) node {$z_{r}$};

	\node[nodop] at (1.5,2.2) {$p_0$};
	\node[nodop] at (2.5,2.2) {$p_1$};
	\node[nodop] at (4.5,2.2) {$p_{r-1}$};
	\node[nodop] at (5.5,2.2) {$p_r$};

\end{tikzpicture}} \caption{Partitions of the network configuration space}\label{fig:defins}
\end{figure}
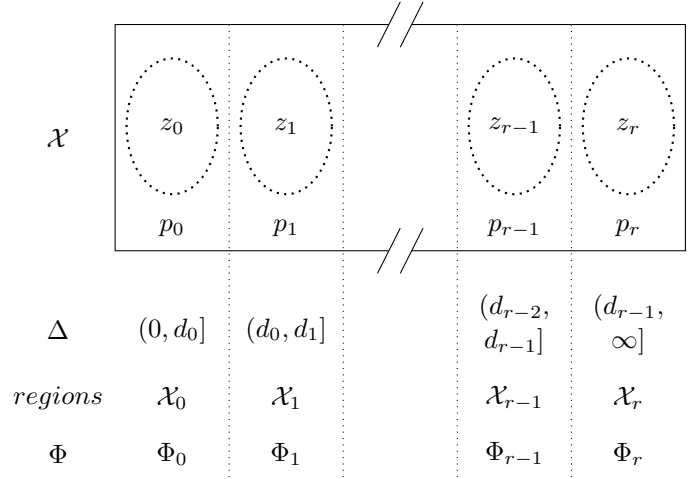

Figure~\ref{fig:defins} illustrates the model and notations used. Given $r$ integers $0<d_0<\dots <d_{r-1}$, let $\Delta=\{\Delta_0\cup \Delta_1\cup \dots \cup \Delta_r\}$ be a set of intervals where $\Delta_0=(0,d_0]$, $\Delta_i=(d_{i-1},d_i]$ for $i\in [1,r-1]$ and $\Delta_r=(d_{r-1},\infty]$. Let ${\cal X}$ be partitioned (and its components called \emph{regions}) as ${\cal X}={\cal X}_0 \cup \dots \cup {\cal X}_r$ where ${\cal X}_i=\{x\in {\cal X} \mid \Delta(x)\in \Delta_i\}$ for $i\in [0,r]$. Let $p_i$ denote the probability that $x$ is one of ${\cal X}_i$. Let also $z_i$ denote the probability that $x$ is one of ${\cal Z}_i$, being ${\cal Z}_i \subseteq {\cal X}_i$ defined in Section~\ref{s:bounded}. Similarly, ${\cal Q}=\{\Phi_0,\Phi_1,\dots,\Phi_r\}$ and $\Phi: {\cal X}\to {\cal Q}$ gets defined by $\Phi(x)=\Phi_i \iff \Delta(x)\in \Delta_i(x)$. 

\section{Crude Monte Carlo Method} \label{s:crude}

A crude Monte Carlo simulation estimates the expected value $\bar{\Phi}$ of $\Phi$ by independently sampling $N$ network configurations $x^{(1)},\dots,x^{(N)}$, computing $\Phi$ for each one, and building an estimator

\begin{equation*}
\widehat{\Phi_N}=\frac{1}{N}\sum_{i=1..N}\Phi(x^{(i)})
\end{equation*}

whose variance is ($EV(\cdot)$ denotes the expected value)

\begin{equation*}
\widehat{\sigma^2_N} = \frac{N}{N^2}\mathit{var}(\Phi) = \frac{1}{N}(EV(\Phi^2)-EV^2(\Phi)) = \frac{1}{N}(\sum_{i=1..N}\Phi_i^2p_i-\bar{\Phi}^2) .
\end{equation*}

Sampling each $x^{(i)}$ involves $m$ Bernoulli trials for determining the state of each edge. Computing $\Phi$ for $x^{(i)}$ is done by applying a breadth-first-search (BFS) algorithm starting at every node of $K$. So each iteration takes a time that is $O(\max\{m,|K|^2\})$.

\section{Proposed Monte Carlo Method} \label{s:bounded}

Suppose that certain topological knowledge about $G=(V,E)$ is available in the form of certain edge sets, called \emph{pathsets}, \emph{cutsets}, \emph{$d_i$-pathsets} and \emph{$d_i$-cutsets} for a given $i\in [0..r-1]$, that we define next. Let $P$ be any subset of $E$ and $G'=(V,P)$ the partial graph of $G$ yielded by $P$. Then,

\begin{itemize}
\item $P$ is a \emph{pathset} if and only if $K$ is connected in $G'$;
\item $P$ is a \emph{\mbox{$d$-pathset}} if and only if $K$ is \mbox{$d$-connected} in $G'$;
\item a \mbox{($d$-)pathset} is said to \emph{operate} when all of its edges operate.
\end{itemize}

Similarly let $C$ be any subset of $E$ and $G'=(V,E\setminus C)$ the partial graph of $G$ yielded by $E\setminus C$. Then,

\begin{itemize}
\item $C$ is a \emph{cutset} if and only if $K$ is not connected in $G'$;
\item $C$ is a \emph{\mbox{$d$-cutset}} if and only if $K$ is not \mbox{$d$-connected} in $G'$;
\item a \mbox{($d$-)cutset} is said to \emph{fail} when all of its edges are failed.
\end{itemize}

Hereafter, unless otherwise specified or clear by the context, the terms pathset and cutset refer to any of the above defined sets, regardless of the presence or absence of a length constraint. In our context, an elementary event is a network configuration $X$, with operational edges $\{e\in E/ X_e=1\}$ and failed edges $\{e\in E/ X_e=0\}$. The sets of operating/failed edges define whether a given pathset/cutset operates/fails. \\
 
Under certain circumstances, the simultaneous occurrence of operating/failed sets allows to know the value of $\Phi$ for a given network configuration. For example, suppose that a certain network configuration $X$ is such that a given \mbox{$5$-pathset} operates while a given \mbox{$2$-cutset} fails. It follows that the maximal distance between the nodes of $K$ must be any of $\{3,4,5\}$ in the partial graph encoded by $X$. If the interval $(2,5]$ belongs to $\Delta$, then the region to which $X$ belongs is known, and so is its $\Phi$ value. The proposed method takes advantage of this property as we see next. Assume that the following sets of edges are known:

\begin{itemize}
\item ${\cal P}_0$: set of some pathsets;
\item ${\cal P}_1,\dots,{\cal P}_{r-1}$: $r-1$ sets such that each ${\cal P}_i$ is a set of some \mbox{$d_i$-pathsets};
\item ${\cal P}_r=\emptyset$ (for convenience of notation);
\item ${\cal C}_0=\emptyset$ (for convenience of notation);
\item ${\cal C}_1,\dots,{\cal C}_{r-1}$: $r-1$ sets such that each ${\cal C}_i$ is a set of some \mbox{$d_{i-1}$-cutsets};
\item ${\cal C}_r$: set of some cutsets.
\end{itemize}

In the previous definitions, the word `some' means that every set can contain any number of elements ranging from zero to the maximum existing number of \mbox{$(d_i$-)pathsets} or \mbox{$(d_i$)-cutsets}. At least one of the sets must be non-empty for the method to be useful; if all sets are empty then the method coincides with crude Monte Carlo. Now, the following events can be defined over $\cal X$:

\begin{align*}
{\cal T}_i& = (\text{some element of }{\cal P}_i\text{ operates}) & (i=0,\cdots,r-1) \\
{\cal K}_j& = (\text{some element of }{\cal C}_j\text{ fails}) & (j=1,\cdots,r) \\
{\cal Z}_0& = {\cal T}_0 \\
{\cal Z}_h& = {\cal T}_h \wedge {\cal K}_h & (h=1,\cdots,r-1)\\
{\cal Z}_r& = {\cal K}_r
\end{align*}

With these definitions, given a network configuration $x$, it holds that $({\cal Z}_0 \to (\Delta(x)\in (0,d_0]))$ and that $({\cal Z}_i \to (\Delta(x)\in (d_{i-1},d_i]))$ for $i\in [1\dots r]$. In other words, each event ${\cal Z}_{h=0\cdots r}$ determines a precise $\Phi$ value. The events ${\cal Z}_0,\dots,{\cal Z}_r$ are subsets of ${\cal X}_0,\dots,{\cal X}_r$ respectively and thus pairwise disjoint too. This is also shown in Figure~\ref{fig:defins}, where $p_i=\Pr(x \in {\cal X}_i)$ and $z_i=\Pr(x \in {\cal Z}_i)$ for $i\in [0\dots r]$.\\

In what follows all summations have an implicit subscript $i=0,\dots,r$. Suppose that the probabilities $z_0,\dots,z_r$ are easy to compute. Then it is easy to compute $\phi=\sum\Phi_iz_i$,  the part of $\bar\Phi$ for which ${\cal Z}$ accounts. The method described below is based on computing the remaining part of $\bar\Phi$ (the one given by the events out of ${\cal Z}$) by restricting the sampling space to ${\cal X}{\setminus}{\cal Z}$. Let $z=\Pr(x\in {\cal Z})=\sum z_i$. Our sampling plan estimates $\bar{\Phi}$ by sampling $N$ network configurations $x^{(1)},\dots,x^{(N)}$ within ${\cal X}{\setminus}{\cal Z}$ (with a probability distribution that respects the relative probabilities among the network configurations in ${\cal X}{\setminus}{\cal Z}$), computing $\Phi$ for each one, and building an estimator $\widetilde{\Phi_N}$ with variance $\widetilde{\sigma^2_N}$ as follows:
\begin{equation*} \widetilde{\Phi_N} = \frac{1}{N}\left(\sum_i\Phi(x^{(i)})(1-z)\right)+\phi \end{equation*}
whose variance is

\begin{equation*} \label{eq:variance-bmc} \begin{split} \widetilde{\sigma^2_N} & = \\
	& \frac{N}{N^2}var\left((1-z)\Phi\right) = \frac{(1-z)^2}{N} \left(EV(\Phi^2)-EV^2(\Phi)\right) =  \\
	& \frac{(1-z)^2}{N} \left( \sum_i\frac{\Phi_i^2(p_i-z_i)}{1-z} - \left( \sum_i\frac{\Phi_i(p_i-z_i)}{1-z} \right)^2 \right) =  \\
	& \frac{1}{N}\left((1-z)\sum_i\Phi_i^2(p_i-z_i)-(\bar{\Phi}-\phi)^2\right).
\end{split} \end{equation*}

Here the expected values involve probabilities conditioned to ${\cal X}{\setminus}{\cal Z}$, hence the application of the correction factor $1/(1-z)$ to $p_i-z_i$.

\subsection{Variance Reduction}

The variances obtained through the crude and the proposed Monte Carlo methods are next compared; for simplicity it is done for one single iteration (i.e. the simulation sample size is one). Single-index summations are over $i \in \{0,\dots,r\}$ and double-index summations over all pairs $(i,j) \in \{0,\dots,r\}^2$.

\begin{equation*} \label{eq:varred} \begin{split}
\widehat{\sigma^2_1} & - \widetilde{\sigma^2_1} = \\
& \sum_i\Phi_i^2p_i + (1-z)\sum_i\Phi_i^2(p_i-z_i) - 2\bar{\Phi}\phi+\phi^2 = \\
& z\sum_i\Phi_i^2p_i-z\sum_i\Phi_i^2z_i+\sum_i\Phi_i^2z_i - \\
	& \qquad 2(\sum_i\Phi_ip_i)(\sum_i\Phi_iz_i) + (\sum_i\Phi_iz_i)^2 = \\
& \sum_{ij}\Phi_i^2p_iz_j - \sum_{ij}\Phi_i^2z_iz_j + \sum_{ij}\Phi_j^2p_iz_j - \\ 
	& \qquad 2\sum{ij}\Phi_i\Phi_jp_iz_j + \sum{ij}\Phi_i\Phi_jz_iz_j = \\
& \sum_{ij}(\Phi_i-\Phi_j)^2p_iz_j + \sum_{ij}(\Phi_i\Phi_j-\Phi_i^2)z_iz_j = \\
& \sum_{ij}(\Phi_i-\Phi_j)^2(p_i-z_i)z_j + \\
	& \qquad \sum_{ij}\left((\Phi_i-\Phi_j)^2+\Phi_i\Phi_j-\Phi_i^2\right)z_iz_j = \\
& \sum_{ij}(\Phi_i-\Phi_j)^2(p_i-z_i)z_j + \sum_{ij}(\Phi_j(\Phi_j-\Phi_i))z_iz_j = \\
& \sum_{ij}(\Phi_i-\Phi_j)^2(p_i-z_i)z_j + \sum_{i<j}(\Phi_j-\Phi_i)^2z_iz_j.
\end{split} \end{equation*}

The following Lemma characterises the variance reduction in terms of the regions and their respective subsets determined by the set of pathsets and cutsets.

\begin{lemma} The difference of the variances $\widehat{\sigma^2_1} - \widetilde{\sigma^2_1}$ is always non-negative. Regarding strict positivity, it is a necessary and sufficient condition that two non-empty regions ${\cal X}_i$ and ${\cal X}_j$ exist such that their $\Phi$ values are different and ${\cal Z}_j$ is non-empty.
\end{lemma} \label{lem:varred}
\bproof
Observe that both summations in the final form of expression of $\widehat{\sigma^2_1} - \widetilde{\sigma^2_1}$ only involve non-negative terms, hence the difference of variances is always non-negative. Regarding strict positivity, assume that two non-empty regions ${\cal X}_i$ and ${\cal X}_j$ exist such that $\Phi_i \neq \Phi_j$, $p_i>0$ and $z_j>0$ (therefore $p_j>0$). Then, if $p_i>z_i$, the first summation will have a strictly positive term given by the subindices $i,j$. If $p_i=z_i$, then $z_i>0$ and therefore the second summation will have a strictly positive term given by the  subindices $i,j$. This proves the sufficiency of the statement about ${\cal X}_i$ and ${\cal X}_j$. Conversely, assume that $\widehat{\sigma^2_1} - \widetilde{\sigma^2_1} >0$. Then there must exist $i,j$ such that at least one of the corresponding terms in the first and second summation is strictly positive. If the term for the first summation is strictly positive, then $\Phi_i>\Phi_j$, $p_i>z_i$ and $z_j>0$. Since $(p_i>z_i\rightarrow p_i>0)$ and $(z_j>0\rightarrow p_j>0)$ the statement about ${\cal X}_i$ and ${\cal X}_j$ holds true. If the term for the second summation is strictly positive, then $\Phi_i>\Phi_j$, $z_i>0$ and $z_j>0$. Again, $p_i>0$ and $p_j>0$ and the statement holds true.
\eproof

\subsection{Sampling within ${\cal X}\setminus {\cal Z}$} \label{ss:sampling}

Let $\Omega=\bigcup_{i=0\dots r}({\cal P}_i \cup {\cal C}_i)$ be the set of all edges occurring in at least one pathset or one  cutset under consideration. Note that the event $\cal Z$ is independent from the state of all edges that do not belong to $\Omega$. Then the state of the edges of $E\setminus \Omega$ can be easily sampled with $m-|\Omega|$ Bernoulli trials.
A general sequential procedure to sample the state of the edges of $\Omega$ is the following. Let $(o_1\dots o_{|\Omega|})\in \{0,1\}^{|\Omega|}$ an $|\Omega|$-tuple encoding the sampled states. Assume that $o_1\dots o_r$ have already been sampled (being $r<|\Omega|$) and let $E_r$ be the event in which the first $r$ edges of $\Omega$ have these sampled states respectively. Then the probability that the $r+1$-th edge of $\Omega$ is operational is (knowing that the edges of $\Omega$ must have states such that the network configuration belongs to ${\cal X}\setminus {\cal Z}$):

\begin{equation*}
\Pr(o_{r+1}= 1 \mid {\cal X}\setminus{\cal Z} \wedge E_r) =
	\Pr(o_{r+1}=1) \frac {\Pr({\cal X}\setminus{\cal Z} \mid  E_r\wedge o_{r+1}=1)} {\Pr({\cal X}\setminus{\cal Z} \mid E_r)}
\end{equation*}

where $\Pr(o_{r+1}=1)$ is the reliability of the $r+1$-th edge of $\Omega$. The function that given the reliabilities $\rho_1,\dots,\rho_{|\Omega|}$ of the edges of $\Omega$ returns the probability of the event ${\cal X}\setminus {\cal Z}$, is a polynomial ${\mathbb P}(\rho_1,\dots,\rho_{|\Omega|})$. So, computing $\Pr({\cal X}\setminus{\cal Z} \mid E_r)$ involves replacing $\rho_1,\dots,\rho_r$ by 0 or 1 according to the states sampled for the first $r$ edges and then evaluating $\mathbb P$. Finding (and evaluating) $\mathbb P$ can be very complex when pathsets and cutsets of the same region highly overlap. Observe that $\Pr({\cal X}\setminus {\cal Z})=1-\Pr({\cal Z})= 1-\Pr(\bigwedge_i{\cal Z}_i)=1-\sum_i\Pr({\cal T}_i\wedge {\cal K}_i)$ (recall the pairwise independence of all ${\cal Z}_i$). To compute each $\Pr({\cal T}_i\wedge {\cal K}_i)$, if every edge involved occurs only in one of ${\cal P}_i$ and ${\cal C}_i$ then it is possible to get factorised expressions (see the Appendix in \cite{SartorCOMCOM2012}), and then building and evaluating the polynomial can be done in time $O(|\Omega|)$. The previous considerations about the overlapping of edge sets within the same region also apply to computing $z_0,\cdots,z_r$. \\

For limited cardinalities of $\Omega$ an alternative approach can be used, consisting in precomputing the probability of the occurrence of each of the $2^{|\Omega|}$ possible sub-configurations that exist when only considering the edges of $\Omega$, in $O(2^{|\Omega|})$ time. Then, sampling their states just involves choosing a sub-configuration at random through a cut-point access on a table accumulating the precomputed probabilities (thus in $O(|\Omega|)$ time). This is the fastest way to sample the states for the edges of $\Omega$, but at the expense of the exponential-in-$|\Omega|$ effort for precomputing the table, that can limit its applicability on large networks.

\section{A heuristic for pathset and cutset generation}\label{s:heurisPatCut}

In this section we introduce a heuristic that generates pathsets and cutsets for every region ${\mathcal X}_i$. We develop an algorithm for the two-terminal problem that can be easily generalised to the problem for general sets $K$. For every $i$, we build a set of $d_i$-pathsets and $d_{i-1}$-cutsets, such that no two elements share edges. This set should ideally be the one that maximises the probability $z_i$ that one $d$-pathset operates and one $d$-cutset fails at the same time, to attain the largest variance reduction that is possible. The basic idea consists of a greedy randomised generation of paths with lengths in $(d_{i-1}, d_i]$, followed by a greedy randomised generation of $d_{i-1}$-cutsets, without using any edge included in the former paths. The cycle is repeated several times and the combination of sets that yield the higher probability is finally chosen.

\subsection{Paths generation}

The algorithm shown in Fig.~\ref{algo:generatePath} receives the graph $G$, source and destination nodes $(s,t)$ and the minimum and maximum distances $\ele_1,\ele_2$ that define the region. It returns a random path whose length is in the interval $[\ele_1,\ele_2]$ or the null ($\bot$) element if no such path could be found. The algorithm proceeds with a greedy selection of nodes ({\tt currNode}) that are added to the path under construction {\tt newPath}. It starts by selecting $s$ and ends after reaching $t$ or achieving a point where it is impossible to complete a path with the required constraints. In each iteration, the shortest path ({\tt shP}) between the current node and $t$ is computed. If there is no such path, or its length exceeds the difference between $\ele_2$ and the length of the already built part, the algorithm returns the null element $\bot$. Otherwise, the next edge can be either the first one of {\tt shP}, or another one different from it. When there are more than one feasible edges, the first one of {\tt shP} is chosen with a probability proportional to the ratio between the length of {\tt shP} and the maximum number of edges that can be added to the path under construction without violating the $[\ele_1,\ele_2]$ constraint. The function {\tt rand()} in the algorithm returns a random uniform real number in [0,1). Therefore the algorithm tends to stick to a shortest path when {\tt newPath} has reached a length that leaves few chances to divert. On the contrary, when there is still room for many more edges than the length of a shortest path, the algorithm will tend to choose, at random, other directions to extend {\tt newPath}. In line 11, care is taken to not repeat a node of {\tt shP}, which would result in having cycles within the {\tt newPath}. Lines 13-16 remove the chosen edge from $G$ for future iterations, append it to {\tt newPath} and update its length $\ele$ and {\tt currNode}. Random is therefore introduced in the decision of wheter to make a step in the direction of a shortest path or not, as well as in the selection of the next node, in case it was decided not to follow the shortest path.

\begin{figure}
\textbf{Procedure generatePath$(G, s, t, \ele_1, \ele_2)$}
\begin{center}
\begin{algorithmic}[1]
\STATE $\ele \leftarrow 0$; $\text{currNode} \leftarrow s$; $\text{newPath} \leftarrow \emptyset$
\WHILE {$\text{currNode} \neq t$}
	\STATE $\text{shP} \leftarrow \text{shortestPath}(G, \text{currNode}, t, \text{newPath})$
	\STATE $\text{dist} \leftarrow \text{length}(\text{shP})$
	\IF {$(\text{shP} = \bot) \vee (\ele + \text{dist} > \ele_2)$}
		\RETURN $\bot$
	\ELSE
		\IF {$(\text{currNode has only one neighbour in }G) \vee (\text{rand}() < (\ele+\text{dist}-\ele_1)/(\ele_2-\ele_1))$}
			\STATE $\text{next} \leftarrow \text{neighbour of currNode in shP}$ 
		\ELSE
			\STATE $\text{next} \leftarrow \text{any neighbour of currNode in } G\setminus \text{shp}$
		\ENDIF
		\STATE $\text{remove the edge (currNode,next) from } G$
		\STATE $\text{append the edge (currNode,next) to newPath}$
		\STATE $\text{currNode} \leftarrow \text{next}$
		\STATE $\ele \leftarrow \ele + 1$
	\ENDIF
\ENDWHILE
\RETURN newPath
\end{algorithmic}
\end{center}
\caption{Heuristic algorithm for generating paths with lengths in $[\ele_1,\ele_2]$}
\label{algo:generatePath}
\end{figure}

\subsection{Cutsets generation}

The algorithm for creating an $\ele$-cutset given a certain integer $\ele$ is shown in Fig.~\ref{algo:generateCutset}. It  receives the graph $G$, the source and terminal nodes $(s,t)$, the integer $\ele$ and a set of edges $H$. It starts by building a first-in-first-out queue with all edges of $G\setminus H$ inserted in increasing order of distance to $s$. Random is introduced by shaking these distances prior to insertion. For example, in our tests we swaped each pair of values in $\vec{d}$ with a probability inversely proportional to their difference. The queue is later used to add each edge to the cutset under construction {\tt newCut} in this shaken order; the idea behind this is that dropping edges in the vecinity of $s$ is a good strategy to find low cardinality $\ele$-cutsets for $s,t$. The set $H$ will be used when invoking this algorithm  to avoid using edges already used for other $\ele$-pathsets or $\ele$-cutsets found prior to the one under construction. The {\tt while} loop proceeds adding edges to {\tt newCut} and dropping them from $G$ until one of the following occur: i) the distance between $s$ and $t$ is greater than $\ele$ (so we have an $\ele$-cutset); or ii) the queue is empty (so no $\ele$-cutset exists, which happens if and only if there was a path whose length is not above $\ele$ built exclusively with edges of $G\setminus H$). After the {\tt while} loop, {\tt newCut} is an $\ele$-cutset not necessarily minimal (i.e. some edges can be dropped from it and still be an $\ele$-cutset). The {\tt for} loop builds an $\ele$-cutset {\tt minCut} that is minimal in this sense (although, in general, it will not be a minimum-cardinality $\ele$-cutset).

\begin{figure}
\textbf{Procedure generateCutset$(G, s, t, \ele, H)$}
\begin{center}
\begin{algorithmic}[1]
\STATE $\vec{d} \leftarrow$ vector of distances between $s$ and every node in $G\setminus H$
\STATE randomAlter($\vec{d}$)
\STATE queue $\leftarrow$ all edges of $G{\setminus}H$ inserted in increasing order of their values in $\bar{d}$
\STATE $\text{newCut} \leftarrow \emptyset$; $\text{flag} \leftarrow \text{true}$
\WHILE {flag}
	\STATE $\text{shP} \leftarrow \text{shortestPath}(G, s, t, \emptyset)$
	\IF {$(\text{shP} = \bot) \vee (\text{length(shP)} > \ele)$}
		\STATE $\text{flag} \leftarrow \text{false}$
	\ELSIF {isEmpty(queue)}
		\RETURN $\bot$
	\ELSE
		\STATE $\text{newEdge} \leftarrow \text{pop(queue)}$
		\STATE remove newEdge from $G$
		\STATE add newEdge to newCut
	\ENDIF
\ENDWHILE
\STATE minCut $\leftarrow \emptyset$
\FORALL {$e \in $ newCut}
	\STATE add $e$ to $G$
	\STATE $\text{shP} \leftarrow \text{shortestPath}(G, s,t,\emptyset)$
	\IF {$(\text{shP} \neq \bot) \wedge (\text{length(shP)}\leq \ele)$}
		\STATE remove $e$ from $G$
		\STATE add $e$ to minCut
	\ENDIF
\ENDFOR
\RETURN minCut
\end{algorithmic}
\end{center}
\caption{Heuristic algorithm for generating an $\ele$-cutset}
\label{algo:generateCutset}
\end{figure}

\subsection{The main heuristic}

Our main algorithm iterates until a given amount of time is spent. In each iteration a disjoint set of edges for a certain region ${\mathcal X}_i$ is generated. Each iteration will begin with the generation of one $d_i$-pathset and one $d_{i-1}$-cutset. It then will continue adding more sets, following a certain sequence that defines a ``version'' of the algorithm. Each iteration may not exceed a certain parameter time MAX\_TIME; if it does then it is discarded and a new iteration is run. For each generated sets $P$ and $C$ of $d_i$-pathsets and $d_{i-1}$-cutsets, the probability of the event $z_i$ that they define is computed; the $P$,$C$ pair with the highest $\Pr(z_i)$ is recorded as the algorithm proceeds and returned after timing out. The algorithm shown in Fig.~\ref{algo:mainPathsCuts} corresponds to the version where the sequence pathset-cutset-pathset-cutset is followed in each iteration; we will denote it as PCPC. It illustrates how to invoke the procedures {\tt generatePath} and {\tt generateCutset}. In Section~\ref{ss:squaregrids} we compare the results obtained with seven different versions. The algorithm receives the graph $G$, source and destination nodes ($s,t$) and the range of distances allowed for the zone $[\ele_1,\ele_2]$. It returns the pair (P, C) whose probability was the highest among all the pairs generated. The pair is not necessarily built by two pathsets and two cutsets. Note the way that {\tt generatePath} and {\tt generateCutset} are invoked in lines 13 and 19. Passing $G\setminus \text{P}$ and $\text{P}\cup \text{C}$ allows respectively to obtain disjoint pathsets and cutsets respect to those so far generated in the current iteration. If a subprocedure is invoked MAX\_TIMES times without succeeding to return a pathset or cutset, a new iteration is started. Similar algorithms PP, PPP, ... and CC, CCC, ... are used for the border regions ${\mathcal X}_0$ and ${\mathcal X}_r$, generating only pathsets or only cutsets (with no length constraint).

\begin{figure}
\textbf{Procedure PCPC$(G, s, t, \ele_1, \ele_2)$}
\begin{center}
\begin{algorithmic}[1]

\STATE bestP $\leftarrow\bot$; bestC $\leftarrow\bot$; highestProb $\leftarrow 0$
\WHILE {ellapsed\_time $<$ MAX\_TIME}
	\STATE P $\leftarrow\emptyset$; C $\leftarrow\emptyset$
	\REPEAT
		\STATE p $\leftarrow$ generatePath($G,s,t,l_1,l_2)$
	\UNTIL {(p $\neq \bot$) or MAX\_TRIES attempts were done} 
	\STATE \textbf{if} (p = $\bot$) \textbf{then} continue //\emph{aborts current ``while'' iteration}
	\STATE P $\leftarrow$ P $\cup$ \{p\}
	\REPEAT
		\STATE c $\leftarrow$ generateCutset($G,s,t,l_1-1,\text{P})$
	\UNTIL {(c $\neq \bot$) or MAX\_TRIES attempts were done} 
	\STATE \textbf{if} (c = $\bot$) \textbf{then} continue //\emph{aborts current ``while'' iteration}
	\STATE C $\leftarrow$ C $\cup$ \{c\}
	\IF {$\Pr$(P,C) $>$ highestProb}
		\STATE highestProb $\leftarrow \Pr$(P,C); bestP $\leftarrow$ P; bestC $\leftarrow$ C
	\ENDIF
	\REPEAT
		\STATE p $\leftarrow$ generatePath($G\setminus\text{P},s,t,l_1,l_2)$
	\UNTIL {(p $\neq \bot$) or MAX\_TRIES attempts were done} 
	\STATE \textbf{if} (p = $\bot$) \textbf{then} continue //\emph{aborts current ``while'' iteration}
	\STATE P $\leftarrow$ P $\cup$ \{p\}
	\IF {$\Pr$(P,C) $>$ highestProb}
		\STATE highestProb $\leftarrow \Pr$(P,C); bestP $\leftarrow$ P; bestC $\leftarrow$ C
	\ENDIF
	\REPEAT
		\STATE c $\leftarrow$ generateCutset($G,s,t,l_1-1,\text{P} \cup \text{C})$
	\UNTIL {(c $\neq \bot$) or MAX\_TRIES attempts were done} 
	\STATE \textbf{if} (c = $\bot$) \textbf{then} continue //\emph{aborts current ``while'' iteration}
	\STATE C $\leftarrow$ C $\cup$ \{c\}
	\IF {$\Pr$(P,C) $>$ highestProb}
		\STATE highestProb $\leftarrow \Pr$(P,C); bestP $\leftarrow$ P; bestC $\leftarrow$ C
	\ENDIF
\ENDWHILE
\RETURN bestP, bestC, highestProb
\end{algorithmic}
\end{center}
\caption{Pseudo-code for the main heuristic; version PCPC}
\label{algo:mainPathsCuts}
\end{figure}

%
\section{Numerical examples}\label{s:tests}
%

This section provides numerical examples based on mesh-like topologies. The simulations are inspired on the following situation. There is a contract between a communication network provider and a customer who needs to periodically exchange data between two sites $s$ and $t$. They agree on a scale of fines, to be paid by the provider, according to the number of hops that each packet undergoes. The aim is to estimate the expected value of the fines that will be paid during the contract lifetime. To do so, simulations based on crude Monte Carlo and the proposed method were run and their results compared. The methods were implemented in \texttt{C++} and the tests were run on an Intel Core2 Duo T5450 machine with 2 GB of RAM, executing $10^7$ iterations (sample size).

\subsection{Test case 1 - ANTEL's transport network}

Test case 1 is based on the countrywide transport network topology of ANTEL, the largest telecommunications provider in Uruguay, shown in Figure~\ref{fig:antel-reduc}. Nodes represent the sites whose interfaces perform routing activity, thus adding significant latency. Links represent the existing paths between these sites. Three scenarios are considered, corresponding to interface failure probability values of 0.10, 0.05 and 0.01 respectively. The test illustrates the effect that the rarity of edge failures has on the attained efficiency relative to crude Monte Carlo. Assume that nodes $4$ and $14$ (shown as squares) are to be connected in a context where low latencies are desirable. Table \ref{tab:fines-1} lists the scale of fines payed according to the hop distance between both nodes. Three scales are used, each one corresponding to a certain value of link reliability in the network model ($0.90$, $0.95$ and $0.99$). The simulations will estimate the expected value of the average fine as well as its variance. The fine scales were proportionaly adjusted so that the expected values of the fines to pay were rather similar, by running short simulations. Note how the fines per region must quickly increase to yield the same expected value of fines when link reliabilities become higher, due to the fact that network configurations with high distances, or disconnected, become rarer events. In other words, this means that the provider can agree on paying higher fines per region still facing the same fine expected value, because of improved link reliabilities. Table~\ref{tab:sets-case-1} shows the pathsets and cutsets employed, using the edge labels of Figure~\ref{fig:antel-reduc}.

\input{antel-reduc.tex}

\begin{table}
  \begin{center}
    \small
\begin{tabular}{ | c | c | c | c | c | }
\hline
$\Delta$ & region & $r_e=0.9$ & $r_e=0.95$ & $r_e=0.99$ \\ \hline \hline
up to 5	  & ${\cal X}_0$  & 0 & 0 & 0 \\ \hline
6 to 7	   & ${\cal X}_1$ & 5 & 30 & 1,000 \\ \hline
above 7 & ${\cal X}_2$ & 10 & 60 & 2,000 \\ \hline \hline
disconnected  & ${\cal X}_3$ & 20 & 120 & 4,000 \\ \hline
\end{tabular}
\end{center}
\caption{Test 1: fines per region.}
\label{tab:fines-1}
\end{table}

\begin{table}
\begin{center} \small
\begin{tabular}{ | c | c | c | }
\hline
$reg.$ & ${\cal P}_i$ & ${\cal C}_i$ \\ \hline \hline
${\cal X}_0$ &      \begin{tabular}{c} ${\cal P}_0=\{\{4,5,9,1\},\{11,12,2,8\}\}$ \end{tabular} & ${\cal C}_0=\emptyset$  \\ \hline
${\cal X}_1$ &      \begin{tabular}{c} ${\cal P}_1=\{\{11,12,2,13,14,15\}\}$ \end{tabular} &  ${\cal C}_1=\{\{1,8\}\}$ \\ \hline
${\cal X}_2$ &      \begin{tabular}{c} ${\cal P}_2=\{\{4,17,10,18,19,$ \\ $20,21,22,15\}\}$ \end{tabular} &  ${\cal C}_2=\{\{1,8,13\}\}$ \\ \hline
${\cal X}_3$ &   ${\cal P}_3=\emptyset$  &    \begin{tabular}{c} ${\cal C}_3=\{\{3,4,11\},\{1,8,15\}\}$ \end{tabular} \\ \hline
\end{tabular}
\end{center}
\caption{Test 1: pathsets and cutsets.}
\label{tab:sets-case-1}
\end{table}

Table~\ref{tab:results-case-1} shows $\hat{\Phi}$ and $\tilde{\Phi}$ (the expected values of $\Phi$ estimated by the crude and proposed methods respectively); the estimations obtained for $\widehat{\sigma^2}$ and $\widetilde{\sigma^2}$; and the total times (in seconds). As above mentioned the scale of fines was set up so that the expected values of the fines to pay were approximately the same (they ranged from 0.341855 to 0.361926). Times spent are essentially the same across the three reliability scenarios, with the crude method taking a time approximately 11\% lower than the proposed method. Observe the significant reductions achieved in the variance by the proposed method (13.63, 45.27 and 940.2 for $r_e$ equal to 0.90, 0.95 and 0.99 respectively). An efficiency comparison can be reported via the \emph{relative efficiency}  $RE$, which is a standard ratio employed in simulation literature (see e.g. \cite{Fishman1986}, \cite{rubino2009rare}), defined as follows:

\begin{equation*}
	RE = \frac{\mathit{var}_\mathit{crude}}{\mathit{var}_\mathit{proposed}} \times \frac{\mathit{time}_\mathit{crude}}{\mathit{time}_\mathit{proposed}} \approx \frac{\widehat{\sigma^2}}{\widetilde{\sigma^2}} \times \frac{\mathit{time}_\mathit{crude}}{\mathit{time}_\mathit{proposed}}
\end{equation*}

The $RE$ expresses the attained variance reduction adjusted by the spent-time ratio; in this case, its values are 12.25, 40.27 and 819.4 respectively for $r_e=0.90$, $r_e=0.95$ and $r_e=0.99$.

\begin{table}
\begin{center} \small
\begin{tabular}{ | l | c | c | c | }
\hline
. & Crude &  Proposed &  ratio \\ \hline \hline
$r_e=0.9$ \\ \hline \hline
\quad $\Phi$ &  0.341855  &  0.341856  &  - \\ \hline
\quad $\sigma^2$ & 4.538235$\times 10^{-7}$ & 3.328820$\times 10^{-8}$ & 13.63 \\ \hline
\quad t(s) & 290.9 &  323.7  & 0.8987 \\ \hline
$r_e=0.95$ \\ \hline \hline
\quad $\Phi$ &  0.361926  &  0.361632  &  - \\ \hline
\quad $\sigma^2$ & 2.467931$\times 10^{-6}$ & 5.451282$\times 10^{-8}$ & 45.27 \\ \hline
\quad t(s) & 285.8 &  321.3  & 0.8896 \\ \hline
$r_e=0.99$ \\ \hline \hline
\quad $\Phi$ &  0.338000  &  0.334667  &  - \\ \hline
\quad $\sigma^2$ & 6.018858$\times 10^{-5}$ & 6.401462$\times 10^{-8}$ & 940.2 \\ \hline
\quad t(s) & 280.4 &  321.7  & 0.8715 \\ \hline
\end{tabular}
\end{center}
\caption{Test 1: numerical results.}
\label{tab:results-case-1}
\end{table}

\subsection{Test 2 - Square grids} \label{ss:squaregrids}

This test illustrates the behaviour of seven versions of the heuristic algorithm. One version (that we call PC) always returns one pathset and one cutset. Two versions (PCP, PCC) can return one extra pathset or cutset respectively. Finally, four versions (PCPP, PCPC, PCCP, PCCC) return two, three or four components whose nature corresponds to each letter. Two network topologies were employed: square grids with $8\times 8$ and $15\times 15$ nodes respectively. Table~\ref{tab:testSqGrids} shows the characteristics of the four instances of the problem that were run for each version of the algorithm. Nodes $s$ and $t$ are specified by their ``$x,y$ coordinates'' in the grid (numbered from zero). The reliability of each edge was randomically set, according to a triangular distribution (0.985, 0.99, 0.995). The parameters MAX\_TIME and MAX\_TRIES were set to 40 seconds and 5 tries. Last column of the table shows the highest probability found among those returned by each version of the algorithm. 

\begin{table}
\begin{center} \small
\begin{tabular}{ | c | c | c | c | c | c | c | }
\hline
Instance & Size & $s$ & $t$ & $\ele_1$ & $\ele_2$ & $\Pr(z_i)$ \\ \hline \hline
Grid1 & $8\times 8$ & (2,2) & (4,4) & 6 & 10 & 1.895E-4 \\ \hline
Grid2 & $8\times 8$ & (2,2) & (5,5) & 8 & 12 & 1.974E-4 \\ \hline
Grid3 & $15\times 15$ & (4,7) & (11,7) & 10 & 20 & 2.529E-6 \\ \hline
Grid4 & $15\times 15$ & (4,4) & (11,11) & 18 & 24 & 9.575E-9 \\ \hline
\end{tabular}
\end{center}
\caption{Test 2: grid-based instances for heuristic tests.}
\label{tab:testSqGrids}
\end{table}

Table~\ref{tab:testSqGridsResults} shows the results of the four tests; within each one, the algorithms are sorted by descending order of the probability that each one returned. Column labelled \%\/best reports the ratio between the returned probability and the highest one for that particular test. Column \#edges reports the total number of edges involved in the pathsets and cutsets of the returned solution. Columns 2s, 3s and 4s report the number of solutions generated that had two, three and four components (pathsets plus cutsets). Finally column best(P,C) reports the number of pathsets and cutsets in the solution returned by each algorithm. First, note that in all cases there is a significant difference between the best and worst returned solutions (their ratio ranging from 1,22 to 2,94). Second, three of the best solutions had four components and the remaining one had three. These results suggest that it might be worth to spend time looking for solutions with many components, rather than striving to get the best ``one pathset - one cutset'' possible solution, in topologies alike. Third, there is no clear winner, although PCCC, PCPC and PCCP seem to have the best overall results.  

\begin{table}
\begin{center} \small
\begin{tabular}{ | c | c | c | c | c | c | c | }
\hline
 & \%\/best & \#edges & 2 sets & 3 sets & 4 sets & best (P,C) \\ \hline \hline
Grid1 \\ \hline \hline
PCCP & 1.000 & 22 & 1,055 & 726 & 687 & 2,2 \\ \hline
PCPC & 0.998 & 24 & 671 & 637 & 255 & 2,2 \\ \hline
PCCC & 0.911 & 18 & 497 & 336 & 71 & 1,3 \\ \hline
PCC & 0.911 & 14 & 1,086 & 729 & 0 & 1,2 \\ \hline
PCP & 0.509 & 18 & 3,186 & 3,052 & 0 & 2,1 \\ \hline
PCPP & 0.501 & 18 & 2,635 & 2,518 & 943 & 2,1 \\ \hline
PC & 0.465 & 10 & 4,135 & 0 & 0 & 1,1 \\ \hline \hline
Grid2 \\ \hline \hline
PCPC & 1.000 & 28 & 417 & 401 & 152 & 2,2 \\ \hline
PCCP & 1.000 & 28 & 604 & 392 & 377 & 2,2 \\ \hline
PCCC & 0.898 & 20 & 463 & 299 & 266 & 1,3 \\ \hline
PCC & 0.898 & 16 & 605 & 387 & 0 & 1,2 \\ \hline
PCP & 0.502 & 22 & 2,012 & 1,907 & 0 & 2,1 \\ \hline
PCPP & 0.501 & 24 & 1,709 & 1,623 & 621 & 2,1 \\ \hline
PC & 0.458 & 12 & 2,439 & 0 & 0 & 1,1 \\ \hline \hline
Grid3 \\ \hline \hline
PCCC & 1.000 & 26 & 82 & 55 & 43 & 1,3 \\ \hline
PCPC & 0.772 & 40 & 79 & 61 & 30 & 2,2 \\ \hline
PCCP & 0.769 & 42 & 105 & 73 & 54 & 2,2 \\ \hline
PCC & 0.667 & 23 & 110 & 74 & 0 & 1,2 \\ \hline
PCPP & 0.394 & 56 & 286 & 196 & 59 & 3,1 \\ \hline
PCP & 0.386 & 37 & 310 & 204 & 0 & 2,1 \\ \hline
PC & 0.340 & 18 & 380 & 0 & 0 & 1,1 \\ \hline \hline
Grid4 \\ \hline \hline
PCP & 1.000 & 50 & 169 & 105 & 0 & 2,1 \\ \hline
PCPP & 0.996 & 52 & 180 & 98 & 64 & 2,1 \\ \hline
PCPC & 0.996 & 52 & 56 & 27 & 13 & 2,1 \\ \hline
PCCP & 0.988 & 62 & 31 & 21 & 12 & 2,2 \\ \hline
PC & 0.821 & 28 & 202 & 0 & 0 & 1,1 \\ \hline
PCC & 0.821 & 28 & 37 & 17 & 0 & 1,1 \\ \hline
PCCC & 0.821 & 28 & 28 & 17 & 12 & 1,1 \\ \hline
\end{tabular}
\end{center}
\caption{Test 2: ranking of results for the grid-based instances.}
\label{tab:testSqGridsResults}
\end{table}

\subsection{Test 3 - Randomised extension of Arpanet}

This test illustrates the combined effect of the heuristic algorithm for pathset and cutset generation and the variance-reduction technique. The network, shown in Fig~\ref{fig:arpanetRandomExt}, has 60 nodes and 110 edges. It was generated by growing the original Arpanet with 40 nodes according to the random network model of \cite{AlbertBarabasi2002}. As in Test 1, three instances were run, each one with all edges set to the same reliability. In this case the nodes $s,t$ are represented as squares. Table~\ref{tab:fines-3} lists the scale of fines, split in three zones; again they were adjusted so that the expected value was similar for the different edge reliabilities. The heuristic PCCP was applied for ${\mathcal X}_1$ and it returned a solution built by one 12-pathset and two 5-cutsets. Heuristics PPPP and CCCC where applied to define the zones ${\mathcal X}_0$ and ${\mathcal X}_2$, returning respectively three pathsets and two cutsets. The parameter MAX\_TIME was set again to 40 seconds. The results of the test are summarised in Table~\ref{tab:results-case-3}; each time reported for the proposed method include the 120 seconds spent generating the pathsets and cutsets. Note the significant relative efficiencies, in particular for rarer failures, obtained in this test: 21.92, 136.15 and 12,615.39 respectively for $r_e$ equal to 0.90, 0.95 and 0.99.  

\begin{table}
\begin{center} \small
\begin{tabular}{ | c | c | c | c | c | }
\hline
$\Delta$ & region & $r_e=0.90$ & $r_e=0.95$ & $r_e=0.99$ \\ \hline \hline
up to 5	  & ${\cal X}_0$  & 0 & 0 & 0 \\ \hline
6 to 12	   & ${\cal X}_1$ & 10 & 150 & 85,000 \\ \hline
above 12 & ${\cal X}_2$ & 1,000 & 15,000 & 8,500,000 \\ \hline
\end{tabular}
\end{center}
\caption{Test 3: fines per region.}
\label{tab:fines-3}
\end{table}

\begin{table}
\begin{center} \small
\begin{tabular}{ | l | c | c | c | }
\hline
. & Crude &  Proposed &  ratio \\ \hline \hline
$r_e=0.90$ \\ \hline \hline
\quad $\Phi$ &  0.928540  &  0.928373  &  - \\ \hline
\quad $\sigma^2$ & 9.135802$\times 10^{-5}$ & 2.591886$\times 10^{-6}$ & 35.25 \\ \hline
\quad t(s) & 606.4 &  975.2  & 0.6218 \\ \hline
$r_e=0.95$ \\ \hline \hline
\quad $\Phi$ &  0.933135  &  0.915608  &  - \\ \hline
\quad $\sigma^2$ & 1.366002$\times 10^{-3}$ & 6.188060$\times 10^{-6}$ & 220.75 \\ \hline
\quad t(s) & 586.2 &  950.4  & 0.6168 \\ \hline
$r_e=0.99$ \\ \hline \hline
\quad $\Phi$ &  0.935000  &  0.928744  &  - \\ \hline
\quad $\sigma^2$ & 7.232224$\times 10^{-1}$ & 3.217148$\times 10^{-5}$ & 22,480.23 \\ \hline
\quad t(s) & 513.1 &  914.4  & 0.5612 \\ \hline
\end{tabular}
\end{center}
\caption{Test 3: numerical results.}
\label{tab:results-case-3}
\end{table}

\input{aleat60-110.tex}

\section{Conclusions and future work}\label{s:conclusions}

The proposed simulation method showed its capability to achieve significant variance reductions when applied on mesh-like networks. The precise conditions under which there is a reduction in the variance of the estimated parameter, with respect to the crude method, were shown in  Lemma~\ref{lem:varred}. The tests also showed that the proposed heuristic was able to generate sets that exploited the mentioned variance-reduction potential at significant levels. Even considering the extra time required for running the heuristic and for sampling with the proposed plan, the efficiency gains are noteworthy, specially when the links become more reliable.

The heuristics hereby introduced, yet resulting in important efficiency gains when chained to the simulation in these simple versions, can be improved in several ways. The algorithm could adapt the amount of effort spent in searching higher cardinality sets of pathsets/cutsets according to statistics on the sets so far found or on the connectivity level of the terminal set. It could also alter the relative effort devoted to the generation of different sequences of pathsets and cutsets, reacting to the results so far obtained during execution. The reliability of each edge should be taken into account when choosing which one to add to the pathset or cutset under construction, particularly in networks where the reliabilities significantly differ. The time spent generating the sets MAX\_TIME could also be initially set according to the sample size and the number of edges and regions (all of which determine at a large extent the time spent by the simulation). Moreover, it could be adjusted during the algorithm execution in light of the number and quality of the so far found sets.

\bibliographystyle{plain}
\bibliography{./biblioDCR}

\tableofcontents

\end{document}

%% file: antel-reduc.tex
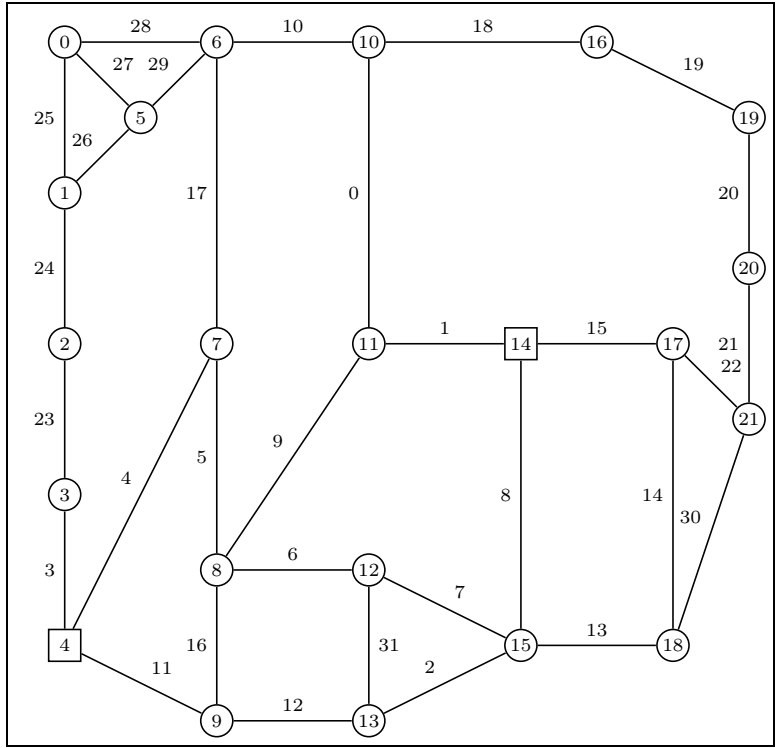
\begin{figure*}[h!]\centering\fbox{
 \begin{tikzpicture}  [scale=1.0,>=stealth,shorten >=0.1pt, auto, semithick,
                     nodop/.style={circle,draw=black,minimum size=12pt, inner sep=0pt, font=\scriptsize},
                     nodot/.style={rectangle,draw=black,minimum size=12pt, inner sep=0pt, font=\scriptsize},
                     every to/.style={draw,thin,black}]
                     
  \begin{scope}[xshift=0cm,scale=1]
\node [nodot] (t4) at (1,1) {4};
\node [nodop] (t3) at (1,3) {3};
\node [nodop] (t2) at (1,5) {2};
\node [nodop] (t1) at (1,7) {1};
\node [nodop] (t0) at (1,9) {0};
\node [nodop] (t5) at (2,8) {5};
\node [nodop] (t9) at (3,0) {9};
\node [nodop] (t8) at (3,2) {8};
\node [nodop] (t7) at (3,5) {7};
\node [nodop] (t6) at (3,9) {6};
\node [nodop] (t13) at (5,0) {13};
\node [nodop] (t12) at (5,2) {12};
\node [nodop] (t11) at (5,5) {11};
\node [nodop] (t10) at (5,9) {10};
\node [nodop] (t15) at (7,1) {15};
\node [nodot] (t14) at (7,5) {14};
\node [nodop] (t16) at (8,9) {16};
\node [nodop] (t18) at (9,1) {18};
\node [nodop] (t17) at (9,5) {17};
\node [nodop] (t21) at (10,4) {21};
\node [nodop] (t20) at (10,6) {20};
\node [nodop] (t19) at (10,8) {19};

\path (t4) edge node {\scriptsize 3} (t3);
\path (t3) edge node {\scriptsize 23} (t2);
\path (t2) edge node {\scriptsize 24} (t1);
\path (t1) edge node {\scriptsize 25} (t0);
\path (t4) edge node {\scriptsize 11} (t9);
\path (t4) edge node {\scriptsize 4} (t7);
\path (t1) edge node {\scriptsize 26} (t5);
\path (t0) edge node {\scriptsize 27} (t5);
\path (t0) edge node {\scriptsize 28} (t6);
\path (t5) edge node {\scriptsize 29} (t6);
\path (t9) edge node {\scriptsize 16} (t8);
\path (t8) edge node {\scriptsize 5} (t7);
\path (t7) edge node {\scriptsize 17} (t6);
\path (t9) edge node {\scriptsize 12} (t13);
\path (t8) edge node {\scriptsize 6} (t12);
\path (t8) edge node {\scriptsize 9} (t11);
\path (t6) edge node {\scriptsize 10} (t10);
\path (t12) edge node {\scriptsize 31} (t13);
\path (t11) edge node {\scriptsize 0} (t10);
\path (t13) edge node {\scriptsize 2} (t15);
\path (t12) edge node {\scriptsize 7} (t15);
\path (t11) edge node {\scriptsize 1} (t14);
\path (t10) edge node {\scriptsize 18} (t16);
\path (t15) edge node {\scriptsize 8} (t14);
\path (t15) edge node {\scriptsize 13} (t18);
\path (t14) edge node {\scriptsize 15} (t17);
\path (t18) edge node {\scriptsize 14} (t17);
\path (t16) edge node {\scriptsize 19} (t19);
\path (t18) edge node {\scriptsize 30} (t21);
\path (t17) edge node {\scriptsize 22} (t21);
\path (t21) edge node {\scriptsize 21} (t20);
\path (t20) edge node {\scriptsize 20} (t19);

 \end{scope}

\end{tikzpicture}}\caption{Test 2: reduced transport network topology of ANTEL, Uruguay's national telecommunications provider}\label{fig:antel-reduc}
\end{figure*}

%% file: aleat60-110.tex
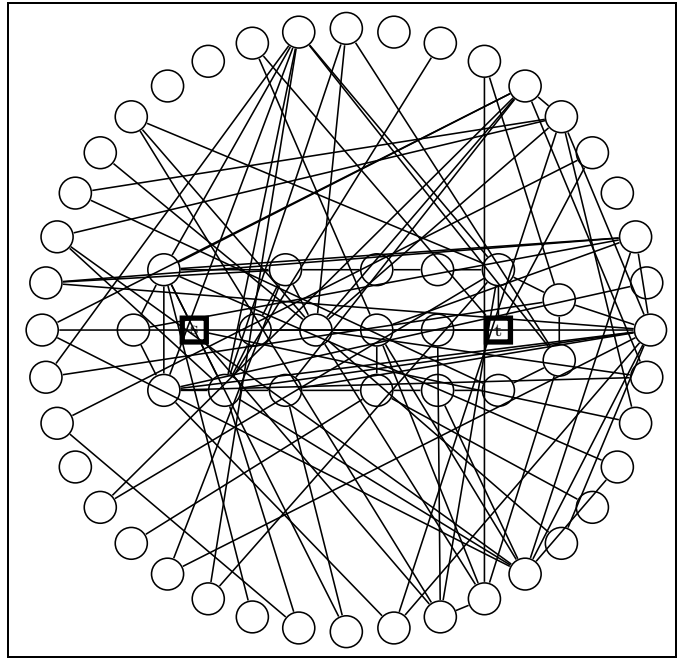
\begin{figure}[h!]\centering\fbox{
 \begin{tikzpicture}  [scale=0.8,>=stealth,shorten >=0.1pt, auto, semithick,
                     nodop/.style={circle,draw=black,minimum size=12pt, inner sep=0pt, font=\scriptsize},
                     nodot/.style={draw=black,line width=0.7mm,minimum size=9pt, inner sep=0pt, font=\scriptsize},
                     every to/.style={draw,thin,black}]
                     
  \begin{scope}[xshift=0cm,scale=1]

\node [nodop] (t0) at (0    , 1) {};
\node [nodop] (t1) at (0.5    , 2) {};
\node [nodop] (t2) at (0.5   ,  0) {};
\node [nodot] (t3) at (1    , 1) {s};
\node [nodop] (t4) at (1.5   ,  0 ) {};
\node [nodop] (t5) at (2 ,  1) {};
\node [nodop] (t6) at (2.5  , 2) {};
\node [nodop] (t7) at (2.5 , 0) {};
\node [nodop] (t8) at (3  , 1) {};
\node [nodop] (t9) at (4  ,  2) {};
\node [nodop] (t10) at (4 , 1) {};
\node [nodop] (t11) at (4  , 0) {};
\node [nodop] (t12) at (5 , 2) {};
\node [nodop] (t13) at (5 , 1) {};
\node [nodop] (t14) at (5  , 0) {};
\node [nodop] (t15) at (6  , 2) {};
\node [nodot] (t16) at (6 , 1) {t};
\node [nodop] (t17) at (6 , 0) {};
\node [nodop] (t18) at (7  ,  0.5) {};
\node [nodop] (t19) at (7  ,  1.5) {};
\node [nodop] (t20) at (8.5,1) {};
\node [nodop] (t21) at (8.43844170297569,1.78217232520115) {};
\node [nodop] (t22) at (8.25528258147577,2.54508497187474) {};
\node [nodop] (t23) at (7.95503262094184,3.26995249869773) {};
\node [nodop] (t24) at (7.54508497187474,3.93892626146237) {};
\node [nodop] (t25) at (7.03553390593274,4.53553390593274) {};
\node [nodop] (t26) at (6.43892626146237,5.04508497187474) {};
\node [nodop] (t27) at (5.76995249869773,5.45503262094184) {};
\node [nodop] (t28) at (5.04508497187474,5.75528258147577) {};
\node [nodop] (t29) at (4.28217232520115,5.93844170297569) {};
\node [nodop] (t30) at (3.5,6) {};
\node [nodop] (t31) at (2.71782767479885,5.93844170297569) {};
\node [nodop] (t32) at (1.95491502812526,5.75528258147577) {};
\node [nodop] (t33) at (1.23004750130227,5.45503262094184) {};
\node [nodop] (t34) at (0.561073738537635,5.04508497187474) {};
\node [nodop] (t35) at (-0.0355339059327373,4.53553390593274) {};
\node [nodop] (t36) at (-0.545084971874736,3.93892626146237) {};
\node [nodop] (t37) at (-0.955032620941839,3.26995249869773) {};
\node [nodop] (t38) at (-1.25528258147577,2.54508497187474) {};
\node [nodop] (t39) at (-1.43844170297569,1.78217232520115) {};
\node [nodop] (t40) at (-1.5,1) {};
\node [nodop] (t41) at (-1.43844170297569,0.217827674798846) {};
\node [nodop] (t42) at (-1.25528258147577,-0.545084971874739) {};
\node [nodop] (t43) at (-0.95503262094184,-1.26995249869773) {};
\node [nodop] (t44) at (-0.545084971874738,-1.93892626146237) {};
\node [nodop] (t45) at (-0.0355339059327386,-2.53553390593274) {};
\node [nodop] (t46) at (0.561073738537634,-3.04508497187474) {};
\node [nodop] (t47) at (1.23004750130227,-3.45503262094184) {};
\node [nodop] (t48) at (1.95491502812526,-3.75528258147577) {};
\node [nodop] (t49) at (2.71782767479885,-3.93844170297569) {};
\node [nodop] (t50) at (3.5,-4) {};
\node [nodop] (t51) at (4.28217232520115,-3.93844170297569) {};
\node [nodop] (t52) at (5.04508497187474,-3.75528258147577) {};
\node [nodop] (t53) at (5.76995249869773,-3.45503262094184) {};
\node [nodop] (t54) at (6.43892626146237,-3.04508497187474) {};
\node [nodop] (t55) at (7.03553390593274,-2.53553390593274) {};
\node [nodop] (t56) at (7.54508497187474,-1.93892626146237) {};
\node [nodop] (t57) at (7.95503262094184,-1.26995249869773) {};
\node [nodop] (t58) at (8.25528258147577,-0.545084971874738) {};
\node [nodop] (t59) at (8.43844170297569,0.217827674798844) {};

\path (t0) edge node {} (t1);
\path (t0) edge node {} (t2);
\path (t1) edge node {} (t2);
\path (t1) edge node {} (t3);
\path (t2) edge node {} (t4);
\path (t3) edge node {} (t4);
\path (t4) edge node {} (t5);
\path (t1) edge node {} (t6);
\path (t5) edge node {} (t6);
\path (t4) edge node {} (t7);
\path (t6) edge node {} (t8);
\path (t6) edge node {} (t9);
\path (t8) edge node {} (t10);
\path (t7) edge node {} (t11);
\path (t10) edge node {} (t11);
\path (t9) edge node {} (t12);
\path (t10) edge node {} (t13);
\path (t11) edge node {} (t14);
\path (t12) edge node {} (t15);
\path (t13) edge node {} (t16);
\path (t15) edge node {} (t16);
\path (t14) edge node {} (t17);
\path (t17) edge node {} (t18);
\path (t15) edge node {} (t19);
\path (t18) edge node {} (t19);
\path (t2) edge node {} (t20);
\path (t4) edge node {} (t20);
\path (t5) edge node {} (t20);
\path (t7) edge node {} (t20);
\path (t19) edge node {} (t20);
\path (t2) edge node {} (t21);
\path (t0) edge node {} (t22);
\path (t1) edge node {} (t22);
\path (t10) edge node {} (t22);
\path (t20) edge node {} (t22);
\path (t16) edge node {} (t24);
\path (t8) edge node {} (t25);
\path (t20) edge node {} (t25);
\path (t1) edge node {} (t26);
\path (t7) edge node {} (t26);
\path (t8) edge node {} (t26);
\path (t25) edge node {} (t26);
\path (t19) edge node {} (t27);
\path (t4) edge node {} (t28);
\path (t4) edge node {} (t30);
\path (t8) edge node {} (t30);
\path (t18) edge node {} (t30);
\path (t1) edge node {} (t31);
\path (t2) edge node {} (t31);
\path (t4) edge node {} (t31);
\path (t15) edge node {} (t31);
\path (t18) edge node {} (t31);
\path (t10) edge node {} (t32);
\path (t16) edge node {} (t32);
\path (t8) edge node {} (t35);
\path (t15) edge node {} (t35);
\path (t8) edge node {} (t36);
\path (t17) edge node {} (t37);
\path (t25) edge node {} (t37);
\path (t4) edge node {} (t38);
\path (t25) edge node {} (t38);
\path (t6) edge node {} (t39);
\path (t20) edge node {} (t39);
\path (t22) edge node {} (t39);
\path (t10) edge node {} (t40);
\path (t26) edge node {} (t40);
\path (t19) edge node {} (t41);
\path (t31) edge node {} (t41);
\path (t24) edge node {} (t42);
\path (t15) edge node {} (t44);
\path (t26) edge node {} (t44);
\path (t11) edge node {} (t45);
\path (t6) edge node {} (t46);
\path (t20) edge node {} (t46);
\path (t15) edge node {} (t47);
\path (t31) edge node {} (t47);
\path (t1) edge node {} (t48);
\path (t4) edge node {} (t49);
\path (t42) edge node {} (t49);
\path (t1) edge node {} (t50);
\path (t7) edge node {} (t50);
\path (t2) edge node {} (t51);
\path (t20) edge node {} (t51);
\path (t25) edge node {} (t51);
\path (t3) edge node {} (t52);
\path (t13) edge node {} (t52);
\path (t15) edge node {} (t52);
\path (t35) edge node {} (t52);
\path (t8) edge node {} (t53);
\path (t10) edge node {} (t53);
\path (t18) edge node {} (t53);
\path (t27) edge node {} (t53);
\path (t52) edge node {} (t53);
\path (t4) edge node {} (t54);
\path (t10) edge node {} (t54);
\path (t14) edge node {} (t54);
\path (t20) edge node {} (t54);
\path (t22) edge node {} (t54);
\path (t38) edge node {} (t54);
\path (t40) edge node {} (t54);
\path (t8) edge node {} (t55);
\path (t20) edge node {} (t55);
\path (t11) edge node {} (t56);
\path (t1) edge node {} (t57);
\path (t54) edge node {} (t57);
\path (t3) edge node {} (t58);
\path (t25) edge node {} (t58);
\path (t2) edge node {} (t59);
\path (t8) edge node {} (t59);
\path (t26) edge node {} (t59);

\end{scope}

\end{tikzpicture}}\caption{Network for Test 3 - Random extension of Arpanet}\label{fig:arpanetRandomExt}
\end{figure}

%% file: RR-8267.bbl
\begin{thebibliography}{10}

\bibitem{AlbertBarabasi2002}
R\'eka Albert and Albert-L\'aszl\'o Barab\'asi.
\newblock Statistical mechanics of complex networks.
\newblock {\em Rev. Mod. Phys.}, 74:47--97, Jan 2002.

\bibitem{Bot-LEcu-Rub-Sim-Tuf-2012}
Z.~I. Botev, P.~L'Ecuyer, G.~Rubino, R.~Simard, and B.~Tuffin.
\newblock Static network reliability estimation via generalized splitting.
\newblock {\em INFORMS Journal on Computing, to appear -}, 2012.

\bibitem{Can-ElKha-Rub-2009}
H.~Cancela, M.~El~Khadiri, and G.~Rubino.
\newblock {\em Rare event analysis by Monte Carlo techniques in static models.
  In Rare Event Simulation using Monte Carlo Methods, G. Rubino, B. Tuffin,
  Eds.}, chapter~7, pages 145--170.
\newblock John Wiley \& Sons, Chichester, 2009.

\bibitem{Can-LEcu-Lee-Rub-Tuf-2009}
H.~Cancela, P.~L'Ecuyer, M.~Lee, G.~Rubino, and B.~Tuffin.
\newblock {\em Analysis and improvements of path-based methods for Monte Carlo
  reliability evaluation of static models. In Simulation Methods for
  Reliability and Availability of Complex Systems, S. M. J. Faulin, A. A. Juan,
  and E. Ramirez-Marquez, Eds.}, pages 65 -- 84.
\newblock Springer-Verlag, Berlin, Germany, 2009.

\bibitem{Can-LEcu-Rub-Tuf-2010}
H.~Cancela, P.~L'Ecuyer, G.~Rubino, and B.~Tuffin.
\newblock Combination of conditional monte carlo and approximate zero-variance
  importance sampling for network reliability estimation.
\newblock In {\em Simulation Conference (WSC), Proceedings of the 2010 Winter},
  pages 1263 --1274, dec. 2010.

\bibitem{SartorCOMCOM2012}
Héctor Cancela, Franco Robledo, Gerardo Rubino, and Pablo Sartor.
\newblock Monte carlo estimation of diameter-constrained network reliability
  conditioned by pathsets and cutsets.
\newblock {\em Computer Communications}, doi=10.1016/j.comcom.2012.08.010,
  2012.

\bibitem{Colbourn87}
Charles~J. Colbourn.
\newblock {\em The Combinatorics of Network Reliability}.
\newblock Oxford University Press, Inc., New York, NY, USA, 1987.

\bibitem{Fishman1986}
George~S. Fishman.
\newblock A {M}onte {C}arlo sampling plan for estimating network reliability.
\newblock {\em Operations Research}, 34(4):581--594, july-august 1986.

\bibitem{Gertsbakh2009}
Ilya~B. Gertsbakh and Yoseph Shpungin.
\newblock {\em Models of Network Reliability: Analysis, Combinatorics, and
  Monte Carlo}.
\newblock CRC Press, Inc., Boca Raton, FL, USA, 1st edition, 2009.

\bibitem{KTI1977}
Hiromitsu Kumamoto, Kazuo Tanaka, and Koichi Inoue.
\newblock Efficient evaluation of system reliability by {M}onte {C}arlo method.
\newblock {\em IEEE Transactions on Reliability}, R-26(5):311 --315, dec. 1977.

\bibitem{LEcu-Rub-Sag-Tuf-2011}
Pierre L'Ecuyer, Gerardo Rubino, Samira Saggadi, and Bruno Tuffin.
\newblock Approximate zero-variance importance sampling for static network
  reliability estimation.
\newblock {\em IEEE Transactions on Reliability}, 60:590--604, 2011.

\bibitem{PR01}
L.~Petingi and J.~Rodriguez.
\newblock Reliability of networks with delay constraints.
\newblock In {\em Congressus Numerantium}, volume 152, pages 117--123, 2001.

\bibitem{Petingi2008}
Louis Petingi.
\newblock A diameter-constrained network reliability model to determine the
  probability that a communication network meets delay constraints.
\newblock {\em WTOC}, 7:574--583, June 2008.

\bibitem{rubino2009rare}
G.~Rubino and B.~Tuffin.
\newblock {\em Rare event simulation using {M}onte {C}arlo methods}.
\newblock Wiley, 2009.

\bibitem{Rub96}
Gerardo Rubino.
\newblock {\em Network reliability evaluation. In State-of-the-art in
  performance modeling and simulation}, chapter~11.
\newblock Gordon and Breach Books, 1996.

\bibitem{Zenklusen-Laumanns-2011}
Rico Zenklusen and Marco Laumanns.
\newblock High-confidence estimation of small s-t reliabilities in directed
  acyclic networks.
\newblock {\em Networks}, 57(4):376--388, 2011.

\end{thebibliography}
